\documentclass[aip,apl,showpacs,twocolumn,floats]{revtex4}
\usepackage{amssymb}

\usepackage{psfig}
\usepackage{dcolumn}
\usepackage{bm}

\begin{document}

\bibliographystyle{prsty}

\title{Voltage from mechanical stress in type-II superconductors: Depinning of the magnetic flux by moving dislocations}
\author{Jaroslav Albert and E. M. Chudnovsky}
\affiliation{\mbox{Department of Physics and Astronomy,} \\
\mbox{Lehman College, City University of New York,} \\ \mbox{250
Bedford Park Boulevard West, Bronx, New York 10468-1589, U.S.A.}
\\ {\today}}

\begin{abstract}\noindent Mechanical stress causes motion of defects
in solids. We show that in a type-II superconductor a moving
dislocation generates a pattern of current that exerts the depinning
force on the surrounding vortex lattice. Concentration of
dislocations and the mechanical stress needed to produce critical
depinning currents are shown to be within practical range. When
external magnetic field and transport current are present this
effect generates voltage across the superconductor. Thus a
superconductor can serve as an electrical sensor of the mechanical
stress.
\end{abstract}

\maketitle

\newpage

Material defects such as dislocations can be set into motion by
subjecting the sample to an external stress. When the stress
becomes large the velocity of dislocations can be as high as the
speed of sound. The dynamics of moving dislocations have been
intensively studied in the past both theoretically
\cite{Alshitz,Mordehai} and experimentally \cite{Nadgornyi}.
Within continuous linear theory of elasticity the dislocation
speed is limited by the shear wave velocity $c_t$ \cite{Hirth}.
When the anharmonicity of the crystal is taken into account the
speed of dislocations has been shown to be intersonic (between
$c_t$ and the speed of longitudinal sound $c_l$) and in some cases
even supersonic \cite{Rosakis, Gumbsch, Grote}, that is above
$c_l$. It has been well established that the fracture of a crystal
under a large external stress is caused by the built-up of
dislocations moving at velocities comparable to the speed of
sound. For this reason, timely detection of fast-moving
dislocations has practical importance for preventing material
fracture. In a transparent material this can be achieved by
optical methods. However, in metals the motion of dislocations is
very difficult to detect. In this Letter we show how this goal can
be achieved in a superconductor.

In type-II superconductors, dislocations have been studied in the
context of vortex pinning \cite{Diaz,Ivlev,Chudnovsky}. The effect
of moving dislocations has not received much attention. A
stationary dislocation is a source of strong pinning provided it
is oriented parallel to the vortex line. Point defects act as weak
pinning sites that may collectively pin vortices in bundles
\cite{Larkin}. The strength of pinning is determined from the
depinning Lorenz force ${\bf F}=(1/c){\bf j}\times {\bf \Phi}_0$
produced by an externally driven critical current $j = j_c$, with
$\Phi_0$ being the flux quantum. At low temperature, when a large
external stress is applied, dislocations accelerate to high
velocities $(v\sim c_t)$ while point defects remain relatively
immobile. It is therefore reasonable to expect that the flux
lattice will be dragged by dislocations in the direction of their
motion. This situation is not generic however because it only
exists when linear dislocations are parallel to the flux lines.
Only in this case the normal core of a vortex line can be
effectively pinned by the dislocation. If the dislocation is at an
angle with the flux line then the pinning is more or less
equivalent to the pinning by a point defect, which is much weaker
than the pinning of the flux line by the entire length of the
dislocation.

In this Letter we examine a more general situation in which
dislocations are not necessarily parallel to the flux lines. The
effect we are going to discuss is not due to pinning of normal
cores of flux lines by the dislocations. We will show that
high-speed dislocations generate superconducting currents of order
$j_c$, thus exerting the depinning force on the surrounding vortex
lattice. Depinning forces produced at a given point in space by an
array of moving dislocations are random. However, at high speed
and sufficient concentration of dislocations these local random
forces will be depinning the entire flux lattice, thus resulting
in a finite resistance of the superconductor. We shall now discuss
the origin of the superconducting current that surrounds a moving
dislocation.

It is well known that a global mechanical rotation of a
superconductor at an angular velocity ${\bf \Omega}$ results in a
macroscopic current. According to the Larmor  theorem, in the
rotating reference frame, Cooper pairs feel the effective magnetic
field ${\bf B} = (2mc/e){\bf \Omega}$, where $e$ and $m$ are the
bare electron charge and mass \cite{Jackson}. This field causes
the Meissner current which is the same in the rotating and
laboratory frames due to the fact that electric current is the
motion of electrons with respect to the ions. Consequently, global
rotation generates the magnetic moment in a superconducting
sample, which is known as the London's effect \cite{London, Alben,
Brickman}. Recently the authors demonstrated \cite{AC} that
high-frequency transverse ultrasound can generate large
superconducting currents via local rotations of the crystal that
occur at an angular velocity \cite{Landau}
\begin{equation}\label{Omega}
{\bf\Omega}({\bf r},t)=\frac{1}{2}\nabla\times\dot{\bf u}({\bf
r},t)\,,
\end{equation}
where ${\bf u}({\bf r},t)$ is the phonon displacement field. In
this Letter we consider a similar effect produced by moving
dislocations. We will show that a moving dislocation is
accompanied by a pattern of the superconducting current. Even far
from the dislocation core this current can exceed $j_c$ when the
dislocation is moving at a high speed. Consequently, an array of
moving dislocations can depin the entire flux lattice.

The electric current is given by
\begin{equation}\label{class-current}
{\bf j}=en_s({\bf v}_s-\dot{\bf u})+en_n({\bf v}_n-\dot{\bf u})\,,
\end{equation}
where $n_s$ and $n_n$ are concentrations of superconducting and
normal electrons, while ${\bf v}_s$ and ${\bf v}_n$ are their
drift velocities respectively. In what follows we will neglect the
contribution of the normal electrons to the total current because
their motion is impeded by viscous forces and is negligible as
compared to the motion of the charged superfluid. For certainty we
will consider deformation ${\bf u}({\bf r},t)$ produced by a
moving screw dislocation. The effect from the edge dislocations is
similar and will be reported elsewhere. Within the continuous
theory of elasticity a screw dislocation along the z-axis, moving
in the x-direction with velocity $v$, is described by the
displacement field \cite{Hirth}
\begin{equation}\label{u}
{\bf u}({\bf r},t)=\frac{bp}{2\pi}\arctan\left(\frac{\gamma
y}{x-vt}\right){\bf e}_z,
\end{equation}
where $b$ is the Burgers vector, $p=\pm1$ is the chirality of the
dislocation, and $\gamma$ is the effective Lorentz factor:
\begin{equation}
\gamma = \left(1 - \frac{v^2}{c_t^2}\right)^{1/2}\,.
\end{equation}
To simplify mathematics, in what follows we will consider the case
of $v^2 \ll c_t^2$. Estimates based upon this approximation will
be valid up to $v \sim 0.3c_t$. We shall see that in fact much
smaller velocities of dislocations may be sufficient to depin the
flux lattice. In this case
\begin{equation}\label{uDot}
\dot{\bf u}({\bf r},t=0)=v \frac{bp}{2\pi}\frac{\sin\theta}{r}{\bf
e}_z\,,
\end{equation}
where $\theta$ is the angle in cylindrical coordinates.

It is convenient to work with the gauge invariant quantity
\begin{equation}
{\bf Q}={\bf A}-(\hbar c/2e){\bm \nabla}\varphi\,,
\end{equation}
where ${\bf A}({\bf r},t)$ is the electromagnetic vector potential
and $\varphi$ is the phase of the superfluid wave function. For
the superconducting current one has
\begin{equation}\label{J}
{\bf j}=-\frac{n_se^2}{mc}\left({\bf Q}+\frac{mc}{e}\dot{\bf
u}\right)\,.
\end{equation}
The term in Eq.\ (\ref{J}) proportional to ${\bf Q}$ is the
standard one and the term proportional to $\dot{\bf u}$ comes from
the motion of the underlying crystal lattice, Eq.\
(\ref{class-current}). The current ${\bf j}$ and the magnetic
field ${\bf B} = {\bm \nabla} \times {\bf A}$ satisfy the Maxwell
equation:
\begin{equation}\label{Maxwell}
{\bm \nabla} \times {\bf B} = \frac{4\pi}{c}{\bf j} +
\frac{1}{c}\dot{\bf E}\,.
\end{equation}
Since ${\bf A}$ produced by a moving dislocation is a function of
$({\bf r}-{\bf v}t)$ and ${\bf E} = -\dot{\bf A}/c$, the last term
in Eq.\ (\ref{Maxwell}) is proportional to $(v/c)^2$ and it can be
safely omitted.  In terms of ${\bf Q}$ Eq.\ (\ref{Maxwell}) then
becomes
\begin{equation}\label{Q}
\lambda^2\nabla\times(\nabla\times{\bf Q})+{\bf
Q}=-\frac{mc}{e}\dot{\bf u}\,,
\end{equation}
where $\lambda=(mc^2/4\pi n_se^2)^{1/2}$ is the London penetration
length.

It is convenient to introduce the dimensionless parameter
$\beta=(bp/2\pi\lambda)$ and dimensionless distance from the
dislocation $\rho=r/\lambda$. Substituting Eq. (\ref{uDot}) into
Eq. (\ref{Q}) we obtain
\begin{equation}\label{Q2}
\rho^2\frac{\partial^2Q_z}{\partial^2\rho}+\frac{\partial^2Q_z}{\partial^2\theta}+
\rho\frac{\partial
Q_z}{\partial\rho}-\rho^2Q_z=\frac{mcv}{e}\beta\rho\sin\theta.
\end{equation}
If we choose solution in the form
$Q_z(\rho,\theta)=(mcv/e)f(\rho)\sin\theta$ then Eq. (\ref{Q2})
becomes an ordinary differential equation for $f(\rho)$:
\begin{equation}\label{f}
\rho^2f''(\rho)+\rho f'(\rho)-(1+\rho^2)f(\rho)=\beta\rho.
\end{equation}
The general solution of this equation that goes to zero at
$\rho\rightarrow\infty$ is
\begin{equation}\label{fSol}
f(\rho)=\beta\left[CK_1(\rho)-\frac{1}{\rho}\right],
\end{equation}
where $K_1(\rho)$ is a modified Bessel function and $C$ is a
constant of integration that can be obtained from the requirement
that ${\bf Q}$ is finite everywhere. Since $K_1(\rho) \rightarrow
1/\rho$ as $\rho \rightarrow 0$, this gives $C=1$. Eq. (\ref{J}) then
gives
\begin{equation}\label{JSol}
{\bf j}=-\frac{c}{4\pi\lambda^2}\left[\frac{mc}{e}\dot{\bf u}+{\bf
Q}\right]= -\frac{mc^2v}{4\pi e\lambda^2}\beta
K_1(\rho)\sin\theta{\bf e}_z.
\end{equation}
Because the angle $\theta$ is defined with respect to the x-axis,
${\bf j}$ vanishes in the plane spanned by the Burgers vector
${\bf b}$ and the dislocation velocity ${\bf v}$. In the yz-plane
the current flows along a closed loop. It generates a dipole-like
magnetic field,
\begin{equation}\label{BSol}
{\bf B}= \nabla\times{\bf Q} = -\frac{mcv}{e\lambda
\rho}[f(\rho)\cos\theta\,{\bf e}_r
-\rho f'(\rho)\sin\theta\,{\bf e}_\theta]
\,.
\end{equation}

The equicurrent lines from an array of moving parallel
dislocations are shown in Fig. 1. As the velocity of dislocations
increases the equicurrent loops in Fig. 1 expand. In the presence
of the transport current, $j_{t}$, normal to a flux line, the line
becomes locally mobile if the combined force exerted on it by
$j_{t}$ and the current $j_d$ due to moving dislocations exceeds
the depinning threshold. To compute this effect one should notice
that the direction and amplitude of $j_d$ fluctuates in space and
time due to random distribution of dislocations. For an ensemble
of parallel dislocations moving at the same speed $v$ the
depinning threshold should be roughly determined by the condition
\begin{equation}\label{depinning-threshold}
{\langle j_d^2 \rangle}^{1/2}\sin\vartheta = {j_c - j_t}\,,
\end{equation}
where $j_c > j_t$ is the critical current in the absence of moving
dislocations and $\vartheta$ is the angle that dislocations make
with the flux lines. The latter enters Eq.\
(\ref{depinning-threshold}) because only the component of ${\bf
j}_d$ normal to the flux line exerts a force on the line.

\begin{figure}
\unitlength1cm
\begin{picture}(18,5.0)
\centerline{\psfig{file=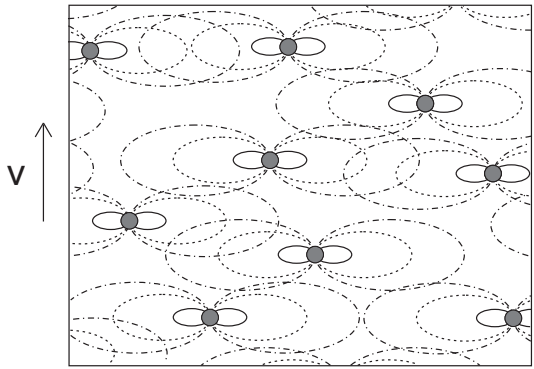,width=7.0cm}}
\end{picture}
\caption{Equicurrent lines for an array of moving parallel
screw dislocations that are normal to the picture.
}
\label{fig2}
\end{figure}

The amplitude of the fluctuating current that appears in Eq.\
(\ref{depinning-threshold}) can be computed as
\begin{equation}
\langle j_d^2 \rangle = n_d\int d^2r {\bf j}^2({\bf r})\,,
\end{equation}
where $n_d$ is a 2D concentration of dislocations and ${\bf
j}({\bf r})$ is given by Eq.\ (\ref{JSol}). At the lower limit
this integral should be cutoff by the size of the dislocation
core, $r \sim b$. This gives
\begin{equation}
\langle j_d^2 \rangle =
n_d\left(\frac{mc^2vb}{8\pi^2e\lambda^2}\right)^2
\pi\ln\frac{\lambda}{b}\,.
\end{equation}
Substituting this result in Eq.\ (\ref{depinning-threshold}), one
obtains the critical (depinning) concentration of dislocations as
function of their velocity:
\begin{equation}\label{final}
\sqrt{n_d} = \frac{\bar{j}_c}{\lambda}\left(1-
\frac{j_t}{j_c}\right)\frac{c_t}{v}\,,
\end{equation}
where we have introduced a dimensionless critical current
\begin{equation}
\bar{j}_c=
\left[\frac{64\pi^{3}}{\ln(\lambda/b)}\right]^{1/2}\frac{
e\lambda^3}{mc^2c_tb\sin\vartheta}\,j_c\,.
\end{equation}

For typical values of the parameters: $\lambda\sim 10^{-5}$cm,
$b\sim 2 \times 10^{-8}$cm, $c_t \sim 2 \times 10^5\,$cm/s, and
$\vartheta = 90^\circ$, the parameter $\bar{j}_c$ is of order
unity at $j_c\sim 10^{5}\,$A/cm$^2$. According to Ref.
\onlinecite{Mordehai} the speed of a screw dislocation very
rapidly approaches the speed of sound on increasing the elastic
stress. Taking $v \sim 0.1c_t$ and $j_t \sim 0.9j_c$ we obtain a
reasonable value of the critical concentration of dislocations:
$n_d \sim 1/\lambda^2$. Even smaller concentration of dislocations
will be required if the transport current is brought closer to
$j_c$. In experiment this effect will manifest itself as a rapid
shift of the critical current towards lower values in the presence
of plastic deformation of the material. The resulting depinning of
flux lines will generate voltage across the superconductor. Since
this voltage originates from the elastic stress, this would be a
remarkable example of a strong non-equilibrium piezoelectric
effect in a conducting material.

This work has been supported by the Department of Energy through
Grant No. DE-FG02-93ER45487.

\end{document}